\documentclass{appolb}
\usepackage{graphicx}

\begin{document}
\title{Dense Hadronic Matter in Neutron Stars%
\thanks{Presented at the LXIII Cracow School of Theoretical Physics Nuclear Matter at Ex-
treme Densities and High Temperatures, Zakopane, Poland, 17–23 September, 2023.}%
}
\author{Laura Tolos
\address{Institute of Space Sciences (ICE, CSIC), Campus UAB, Carrer de Can Magrans, 08193, Barcelona, Spain; \\
Institut d'Estudis Espacials de Catalunya (IEEC), 08034 Barcelona, Spain; \\
Frankfurt Institute for Advanced Studies,  University of Frankfurt, Ruth-Moufang-Str. 1, 60438 Frankfurt am Main, Germany}
}
\maketitle
\begin{abstract}
In this lecture we discuss the properties of dense hadronic matter inside neutron stars. In particular, we pay attention to the role of strangeness in the core of neutron stars, by analysing the presence of baryons and mesons with strangeness. We consider two interesting possible scenarios in their interior, that is, the existence of hyperons leading to the so-called hyperon puzzle and the  presence of a kaon condensed phase inside neutron stars.
\end{abstract}

\section{A short introduction to neutron stars}
  
Neutron stars (NSs) are an excellent laboratory to study the properties of matter under extreme conditions of density, isospin asymmetry and temperature as well as in the presence of strong gravitational and magnetic fields. 

\begin{figure}[htb]
    \centering
    \includegraphics[width=0.7\textwidth]{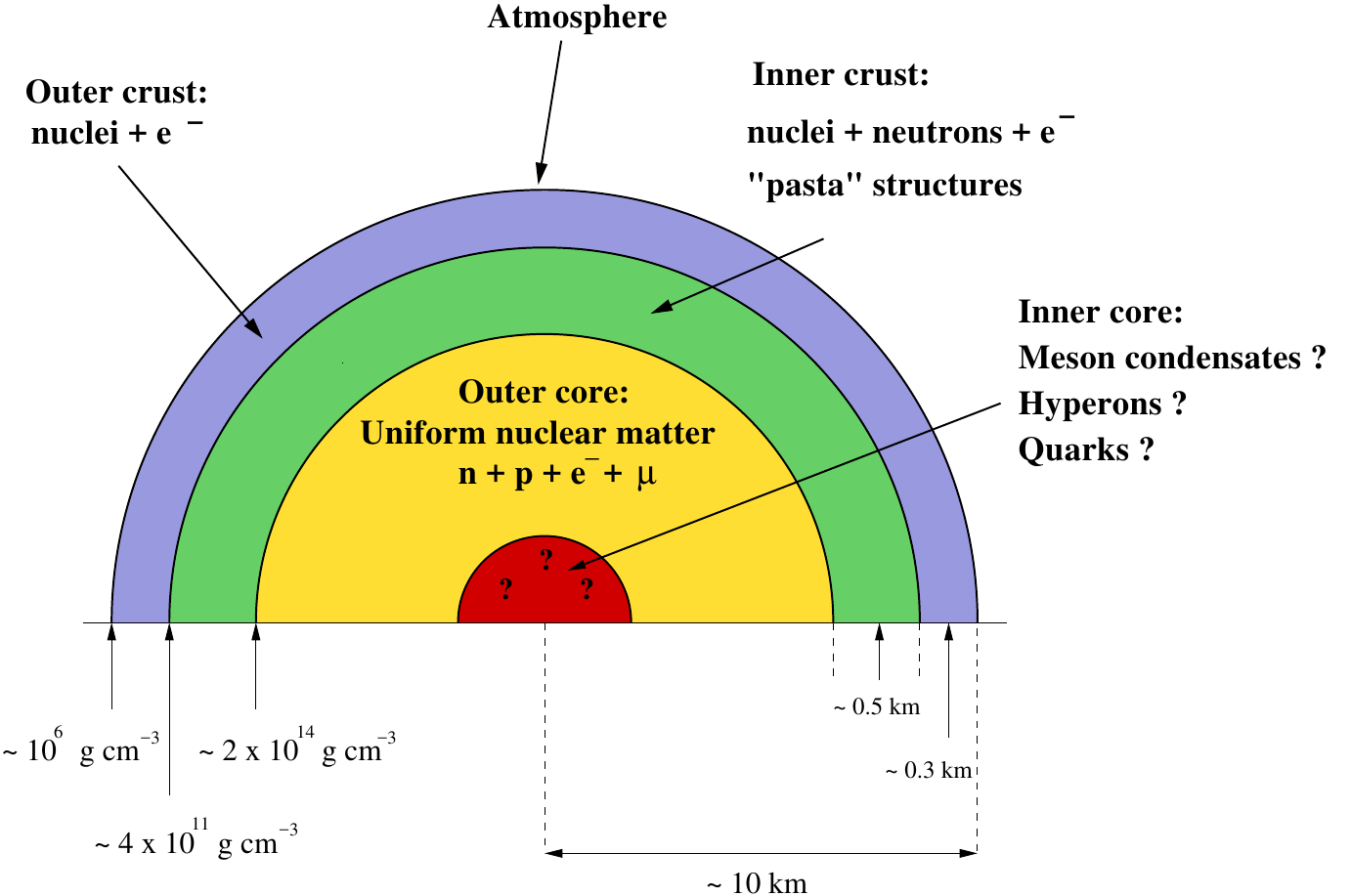}
    \caption{ Illustrative representation of the interior of a neutron star. Figure adapted from Ref.~\cite{Vidana:2018lqp} and taken from Ref.~\cite{Tolos:2020aln}. }
    \label{fig:NSscheme}
\end{figure}

NSs are the final product of core-collapse supernovae. They are in hydrostatic equilibrium with the gravitational collapse mainly counterbalanced by the neutron degeneracy pressure. NSs usually have masses around  1-2~M$_{\odot}$ and radii about 10-12 km, that leads to average densities of $\sim$ $10^{14} {\rm g/cm}^3$,  hence, to very compact stellar objects. However, these stars show an onion-like configuration, where densities extend over a large range. A diagrammatic representation of their internal structure is shown in Fig.~\ref{fig:NSscheme}, where several layers can be seen, that is,  the atmosphere; the outer and inner crust, with $\approx$ 1 km; and the core, splitted in the outer and  inner core, with a radius of $\approx$ 10 km which contains almost the total mass of the NS. 

In spite of the fact that the inner region of an NS is  the largest part and, therefore, the one that determines the properties of an NS, its composition is not known. Thus, several hypothesis have been put forward. These include the presence of matter made of hadrons, such as  baryons or mesons, and/or the existence of deconfined quark matter. 

In this lecture we aim at describing the properties of hadronic matter inside NSs and the consequences for the structure of these compact objects, paying a special attention to the role of strangeness in the interior of NSs. We refer the reader to \cite{Tolos:2020aln,Watts:2016uzu,Burgio:2021vgk} for recent reviews on dense matter in NSs. 

We  start by assuming baryonic matter inside NSs. NSs are charged neutral objects equilibrated by weak interaction processes, that is, they are in $\beta$-equilibrium. This equilibrium can expressed as
\begin{eqnarray}
b_1 \rightarrow b_2+l+\bar{\nu_l}, \hspace{1cm} b_2 + l  \rightarrow b_1 + \nu_l,
\end{eqnarray}
 where $b_i$ refers to a certain type of baryon, $l$ represents a lepton, and $\nu_l$ and $\bar \nu_l$ are the corresponding neutrino and antineutrino, respectively. The composition of the interior of NSs is determined by studying all the possible weak reactions among the different species inside the core. This can be expressed by means of the particle chemical potentials $\mu_i$, such as
\begin{eqnarray}
\mu_{b_i}&=&B_{b_i} \mu_n - q_{b_i} \mu_e , \nonumber \\ 
\mu_{l_j}&=&-q_{l_j} \mu_e ,
\end{eqnarray}
where $B_{b_i}$ is the baryonic number of a given baryon $b_i$, $q_{b_i(l_j)}$ is the charge of $b_i$ baryon ($l_j$ lepton), and $\mu_n$ and $\mu_e$ are the chemical potentials of the neutron and electron, respectively. We note that (anti-)neutrinos freely escape without contributing to the energy balance as their mean-free path is larger than the typical size of an NS.

The charged neutrality is guaranteed by 
\begin{eqnarray}
\sum_{b_i} q_{b_i} \rho_{b_i} + \sum_{l_j} q_{l_j} \rho_{l_j}=0 ,
\end{eqnarray}
where $\rho_{b_i(l_j)}$ is the density of $b_i (l_j)$, with the total baryonic density given by
\begin{eqnarray}
\rho=\sum_{b_i} B_{b_i} \rho_{b_i}.
\end{eqnarray}
 
 In order to connect the microphysics to the bulk properties of an NS, specifically the mass and radius of an NS, we need to solve the so-called structure equations for NSs. These can be determined by means of Einstein's general relativity theory.  In the case of a spherical static star, the Einstein's field equations become the  Tolman-Oppenheimer-Volkoff (TOV) structure equations\footnote{ We note that NSs are usually detected as pulsars, that is, rotating stars. Therefore, the spherical symmetry is broken and  the axial symmetry is the only symmetry remaining as NSs flatten with rotation. This leads to study rotating NSs with a perturbative method developed by Hartle and Thorne \cite{Hartle:1968si}. }:
\begin{eqnarray}
&&\frac{dP(r)}{dr}=-\frac{1}{r^2} \left[\varepsilon (r) + P(r)\right] [M(r)+4 \pi r^3 P(r)] \left[1- \frac{2M(r)}{r} \right]^{-1}, \label{TOVeq1} \\
&& \frac{dM(r)}{dr}=4 \pi  r^2 \varepsilon(r). \label{TOVeq2}
\end{eqnarray}
where we have used $G=c=1$ units.

The TOVs are a set of coupled equations that describe the hydrostatic equilibrium obtained in general relativity. Their interpretation is straightforward. From Eq.~(\ref{TOVeq2}), the mass energy of a shell of matter of radius $r$ and thickness $dr$ can be obtained. As for the left-hand side of Eq.~(\ref{TOVeq1}), this is related to  the net force acting outwards on the surface of the shell by the pressure difference between the interior and the exterior,  $dP(r)$, while the right-hand side of this equation comes from the force of gravity acting on the shell due to the mass accumulated in the interior.

In order to solve the TOVs, we need to determine the pressure $P$ and the associated energy density $\varepsilon$ for a  given composition of the interior of the NS, that is, we need to obtain the so-called equation of state (EoS). Once the EoS is fixed, the TOV equations can be integrated by fixing the initial conditions to the enclosed mass and the pressure at the center of the NS, $M(r=0)=0$ and $P(r=0)=P_c$, with $P_c$ taking an arbitrary value from the EoS. The integration over the radial coordinate $r$ finishes when $P(r=R)= 0$, with $R$ being the radius  and $M(R)$ the total mass $M$ of the star, respectively.

\section{The nuclear equation of state}
\label{sec:nucEoS}

As we mentioned earlier, the gravitational collapse of NS is mainly counterbalanced by the neutron degeneracy pressure. Thus, one of the first hypothesis for baryonic matter would be to consider the interior of NSs as a neutron Fermi gas. However, this is a very unrealistic scenario. On the one hand, an NS must contain a  small fraction of protons and electrons so as to inhibit neutrons from decaying into protons and electrons by their weak interactions. On the other hand,  the Fermi gas model ignores nuclear interactions, which give important contributions to the energy density. Therefore,  we start by assuming that the core of an NS is made of nuclear matter.

The EoS of nuclear matter describes an idealised infinite uniform system made of nucleons (protons and neutrons), where the Coulomb interaction is switched off. Symmetric nuclear matter refers to a system with an equal number of neutrons and protons, and is  the easiest approximation to bulk matter in heavy atomic nuclei.  Pure neutron matter, on the other hand, is the simplest approach to hadronic matter in the NS core.

The energy per nucleon of the nuclear system for a given density $\rho$ can be expressed as
\begin{equation}
    \frac{E}{A}(\rho,\delta) = \frac{E}{A}(\rho,0)+S(\rho) \delta^2 + ...,
\end{equation}
with $\delta=(N-Z)/A$ and $A=N+Z$, being $N(Z)$  the neutron (proton) number. The  energy per nucleon of symmetric nuclear matter ($\delta=0$) is given by $(E/A)(\rho,0)$, while $S(\rho)$ is the symmetry energy that measures the energy cost involved in changing the protons into neutrons. If we expand both terms around nuclear saturation density, $\rho_0$, we obtain
\begin{eqnarray}
\frac{E}{A}(\rho,0)= \frac{E}{A}(\rho_0)+\frac{1}{18}K_0 \epsilon^2 + ... , \nonumber \\
S(\rho)=S_0+\frac{1}{3}L \epsilon + \frac{1}{18} K_{\rm sym} \epsilon^2 +...,
\end{eqnarray}
where $\epsilon=(\rho-\rho_0)/\rho_0$.  The binding energy per nucleon at saturation density, $(E/A)(\rho_0)$, and the incompressibility at the saturation point, $K_0$, are called  isoscalar parameters, whether  the symmetry energy coefficient at saturation density, $S_0$, and $L$ and $K_{\rm sym}$, that give the density dependence of the symmetry energy around saturation, are usually refered as isovector parameters. These  parameters are given by
\begin{eqnarray}
&&K_0 \equiv 9 \rho_0^2 \left( \frac{\partial^2 (E/A)(\rho,\delta)}{\partial \rho^2}\right)_{\rho_0,\delta=0}, \hspace{1cm}
S_0 \equiv \frac{1}{2} \left( \frac{\partial^2 (E/A)(\rho,\delta)}{\partial \delta^2}\right)_{\rho_0,\delta=0}, \nonumber \\
&&L \equiv 3 \rho_0\left( \frac{\partial S(\rho)}{\partial \rho}\right)_{\rho_0}, \hspace{1cm}
K_{\rm sym} \equiv 9 \rho_0^2 \left( \frac{\partial^2 S(\rho)}{\partial \rho^2}\right)_{\rho_0}.
\label{eq:eosparam}
\end{eqnarray}

The  energy density of the system $\varepsilon(\rho,\delta)$  and the pressure $P$ are straightforwardly obtained by
\begin{eqnarray}
&&\varepsilon(\rho,\delta)=\rho \frac{E}{A}(\rho,\delta), \nonumber \\
&&P(\rho,\delta)=\rho^2\frac{\partial (E/A)(\rho,\delta)}{\partial \rho}=\rho \frac{\partial \varepsilon(\rho,\delta)}{\partial \rho}- \varepsilon(\rho,\delta). 
\end{eqnarray}

\subsection{Constraints on the nuclear equation of state}
\label{sec:constraints}

Several constraints on the nuclear EoS can be obtained from nuclear experiments and/or observations. Nonetheless, we have to take several of these constraints with caution since they are determined after using theoretical modelling and/or by means of extrapolations to domains not attainable by experiments and/or observations. In this section we present some experimental and observational constraints that are usually considered for constraining the nuclear EoS.

\subsubsection{Constraints from nuclear experiments}
\label{sec:expconst}

Several constraints on the isoscalar and isovector parameters can be extracted from nuclear experiments.  

With regard to the previously mentioned isoscalar parameters of the nuclear EoS, the values for the nuclear saturation density $\rho_0=0.15-0.16 \ {\rm fm}^{-3}$ and the binding energy per nucleon at that density $(E/A)(\rho_0)=-16 \pm 1$ MeV have been determined from the measurement of density distribution \cite{DeVries:1987atn} and nuclear masses \cite{AUDI2003337}. As for the incompressibility at saturation density $K_0$, the extraction of its value is complicated and results have a wide spread of  $K_0\sim$ 200-300 MeV (see for example \cite{Blaizot:1980tw,Piekarewicz:2003br,Khan:2012ps}). 

As for the isovector parameters $S_0$, $L$ and $K_{\rm sym}$, these can be extracted from experiments involving isospin diffusion measurements \cite{Chen:2004si}, analysis of giant  \cite{Garg:2006vc} and pygmy resonances \cite{Klimkiewicz:2007zz,Carbone:2010az}, isobaric analog states \cite{Danielewicz:2008cm}, isoscaling \cite{Shetty:2007zg}, production of pions \cite{Li:2004cq} and kaons \cite{Fuchs:2005zg,Hartnack:2011cn,Song:2020clw} in heavy-ion collisions (HiCs)  or data on neutron skin thickness of heavy nuclei \cite{Brown:2000pd,Horowitz:2000xj,Horowitz:2001ya,Centelles:2008vu}. Whereas $S_0$ is relatively well constrained around $\sim 30$ MeV, $L$ and $K_{\rm sym}$ are still  poorly known.

\subsubsection{Constraints  from neutron star observations}
\label{sec:obsconstraints}

Other sources to constrain the nuclear EoS come from NS observations, such as masses and radii, and more recently from gravitational wave detection.

The mass of an NS can be determined if the NS is located in a binary system by means of using the Kepler's law modified by general relativity effects. In binary systems, there exist five Keplerian (also called orbital parameters) that can be measured with good precision. These are the binary orbital period ($P_b$), the orbit's eccentricity  ($e$), the projection of the semi-major axis onto the line of sight ($x \equiv a_1 {\rm sin} \, i$, with $i$ being the inclination angle of the orbit), and the time of the periastron ($T_0$) and its longitude ($\omega_0$). Using Kepler's law,  the so-called mass function can be obtained.  This is a relation between the masses of both stars and some of the observed orbital parameters, that is, $f(M_P,M_C,i)=M_c^3 \ {\rm sin^3} i/(M_P + M_C)^2= 4\pi^2 x^3/P_b^2$, with $M_P$ representing the mass of the NS ($P=$ pulsar) and $M_C$ being the mass of its companion. 

In order to determine both masses, we need more information. We then resort to determine the deviations from the Keplerian orbit due to general relativity effects. These effects can be described by the so-called post-Keplerian parameters. These post-Keplerian parameters are the advance of the periastron ($\dot{\omega}$), the changes in the transverse Doppler shift together with the gravitational redshift around an elliptical orbit ($\gamma$),  the range ($r$) and shape ($s$) of the Shapiro time delay, and the orbital decay ($\dot{P_b}$). We note that the post-Keplerian parameters are dependent on the Keplerian parameters and the two masses in the binary. Therefore, if we could determine at least two of them as well as the mass function, we could obtain the masses of the two stars. The further determination of a third post-Keplerian parameter will result in a test of general relativity.

Nowadays more than 2000 pulsars have been discovered, with some of their masses very well determined. The detection and measurement of the masses of the Hulse-Taylor pulsar and its companion \cite{Hulse:1974eb} result in  the Nobel Prize in 1993 for Hulse and Taylor, because it allowed to a test Einstein's general relativity.  More recent accurate values of $2M_{\odot}$ have been  reported, such as for the PSR J1614-2230 \cite{Demorest:2010bx,Fonseca:2016tux}, the PSR J0348+0432 \cite{Antoniadis:2013pzd}, and the  PSR J0740+6620 of $2.14_{-0.09}^{+0.10} M_{\odot}$  \cite{2019NatAs.tmp..439C}. As we will later discuss, these $2M_{\odot}$ measurements are sometimes in tension with theoretical predictions for EoSs that take into account the presence of hyperons.

With regard to radii,  these were extracted from studying the X-ray spectra emitted by the NS atmosphere. This is a rather difficult task as the X-ray spectra strongly depends on the distance to the star, its magnetic field and the composition of the atmosphere.  However, very recently, the situation has dramatically improved with space missions such as 
NICER (Neutron star Interior Composition ExploreR) \cite{2014SPIE.9144E..20A},  and the future STROBE-X (Spectroscopic Time-Resolving Observatory for Broadband Energy X-rays) \cite{strobex} and  eXTP (enhanced X-ray Timing and Polarimetry)  \cite{Watts:2018iom},  since high-precision X-ray astronomy offers precise determinations of masses and radii in a simultaneous way.  Simultaneous measurements of masses and radii are already becoming available from NICER, with the first precise measurements of the radii and masses of the millisecond pulsars PSR J0030+0451 and PSR J0740+6620  \cite{Riley:2019yda,Miller:2019cac,Riley:2021pdl,Miller:2021qha}.

Furthermore, the detection of gravitational waves coming from the merger of two NSs by the LIGO and VIRGO collaborations \cite{TheLIGOScientific:2017qsa,Abbott:2018wiz} has opened new frontiers. Gravitational waves from the late inspirals of NSs depend on the EoS, via the so-called tidal deformability. Indeed, the tidal deformability depends on the NS compactness. Therefore, the measurement of the tidal deformability helps to constrain the EoS.

\subsection{Theoretical models for the nuclear equation of state}

Nuclear matter inside the core of an NS can be described by means of different theoretical many-body approaches, that are usually classified between microscopic ab-initio schemes and phenomenological approachess. 

Microscopic ab-initio approaches refer to schemes where the nuclear EoS is obtained by solving the many-body problem  from two-nucleon (NN) and three-nucleon interactions (NNN). These NN and NNN interactions are fitted to scattering data and finite nuclei. These schemes include the ones based on the variational analysis \cite{Akmal:1998cf}, quantum-montecarlo methods \cite{Wiringa:2000gb,Carlson:2003wm,Gandolfi:2009fj}, the formalism of the correlated basis function  \cite{Fabrocini:1993eaz}, diagrammatic approaches  (among them, the Brueckner-Bethe-Goldstone expansion \cite{Day:1967zza}, the Dirac-Brueckner-Hartree-Fock (DBHF) method \cite{TerHaar:1986xpv,Brockmann:1990cn} and the self-consistent Green's function scheme \cite{2005mbte.book.....D}), renormalization group methods \cite{Bogner:2003wn}, and lattice Quantum Chromodynamics (LQCD) computations \cite{Beane:2010em,Ishii:2006ec}. Whereas the advantage of the vast majority of these approaches is being able to systematically add higher-order contributions that allows for a controlled determination of the nuclear EoS, the main disadvantage lies on the applicability to large densities, since incorporating higher-order terms makes the computations more difficult.

Phenomenological schemes are based on interactions that depend on the density and are adjusted to nuclear observables and observations coming from NSs. Among others, we have non-relativistic energy-density functionals, such as the Skyrme \cite{Skyrme:1959zz} or Gogny \cite{Decharge:1979fa} approaches, or relativistic models, usually derived from an hadronic Lagrangian, using the mean-field or Hartree-Fock approximations \cite{Boguta:1977xi,Serot:1984ey}. The clear advantage of these approaches is the applicability to large densities, while the disadvantage lies on the non-systematic character of these approaches.

For more details on these approaches, we refer to the recent works of Refs.~\cite{Burgio:2021vgk,Oertel:2016bki,Burgio:2018mcr}.

\section{Equation of state with strangeness}
\label{sec:eos-strange}

Given the extreme density conditions  inside NSs as compared to those found on Earth, the existence of new phases of matter in their interior is therefore possible. In particular, the presence of strange baryons (also called hyperons) and strange mesons (antikaons) in the interior of NSs has been explored extensively over the years. In this section we aim at describing strange hadronic matter and the consequences for the structure of NSs. In particular, we will discuss the so-called hyperon puzzle and the phenomenon of kaon condensation in NSs.

\subsection{Strange baryons: Hyperons}
Hyperons are baryons with one or more strange quarks. They are usually denoted by Y and refer to $\Lambda$, $\Sigma$ and $\Xi$. In this part of the lecture we aim at describing the role of hyperons inside NSs. To that end, we first start by summarizing the present experimental status of the YN and YY interactions, followed by shortly describing several theoretical approaches for the YN and YY interactions. We continue by analyzing the properties of hyperons in a many-body baryonic system, and finally address the presence of hyperons in NSs and the hyperon puzzle.

\subsubsection{Experimental status for YN and YY interactions}
\label{exp}

In contrast to the NN system, the YN and YY interactions are poorly constrained. The experimental difficulties arise from the short life of hyperons together with the low-density beam fluxes. Whereas for the $\Lambda$N and $\Sigma$N systems there are hundred of  scattering events \cite{Engelmann:1966npz,Alexander:1968acu,Sechi-Zorn:1968mao,Kadyk:1971tc,Eisele:1971mk}, few exist for $\Xi$N system and non scattering data is available for YY. 

Alternatively, the study of hypernuclei, that is, bound systems composed of nucleons and one or more hyperons can give us information on the YN and YY interactions. More than 40 single $\Lambda$-hypernuclei, and a few double-$\Lambda$ and single-$\Xi$ ones have been detected. As for $\Sigma$ hypernuclei the experimental confirmation is ambiguos, indicating that the $\Sigma$N interaction is most probably repulsive. For a short review on hypernuclei, we refer to Ref.~\cite{Vidana:2018bdi}.

More recently, femtoscopy has emerged as a very interesting tool to study reactions among hadrons \cite{Fabbietti:2020bfg}. Femtoscopy consists in measuring the hadron-hadron correlation in momentum space by obtaining the ratio of the distribution of relative momenta for pairs produced in the same collision and in different collisions (mixed events).  If the measured correlation is larger than one, the interaction is attractive, whereas the values are between zero and one if the interaction is repulsive or a bound state exists. Thus, the comparison of the measured correlation functions with the theoretical predictions will give us information on hadron-hadron interactions, and in particular, on the YN and YY interactions.

\subsubsection{Theoretical approaches to YN and YY interactions}
\label{theory}

In the past  there has been a lot of theoretical progress  trying to describe the YN and YY interactions. The theoretical schemes can be grouped in meson-exchange schemes, chiral effective field theory ($\chi$EFT) approaches, calculations on LQCD, low-momentum schemes and quark-model potentials.

The basic idea in {meson-exchange models} is that the interaction between two baryons is mediated by the exchange of mesons. Starting from the NN meson-exchange model, SU(6)$_{\rm flavor}$ symmetry is assumed to obtain the YN and YY interactions in the  J\"ulich  \cite{Haidenbauer:2005zh} potential, whereas SU(3)$_{\rm flavor}$ symmetry for the Nijmegen \cite{Rijken:2010zzb} ones.

As for the $\chi$EFT schemes, the YN and YY interactions have been built by the J\"ulich-Bonn-Munich group starting from their previous NN $\chi$EFT approach \cite{Polinder:2006zh,Haidenbauer:2013oca,Haidenbauer:2019boi}.

Regards to LQCD, the QCD path integral over the quark and gluon fields at each point of a four-dimensional space-time grid is solved by means of Monte Carlo techniques.   HALQCD \cite{halqcd} and the NPLQCD \cite{nplqcd} collaborations are pioneers in this respect.

And, finally, the YN and YY interactions have been described using other schemes that include {low-momentum interactions} and {quark-model potentials}. The former determines a universal effective low-momentum potential for YN and YY using renormalization-group methods \cite{Schaefer:2005fi}, whereas the latter builds the YN and YY interactions within constituent quark models \cite{Fujiwara:2006yh}.

\subsubsection{Hyperons in dense matter}
\label{dense}

The properties of hyperons in dense matter can be obtained from YN and YY interactions by means of incorporating corrections from the surrounding many-body medium. Within this microscopic formulation, one of the most used scheme is the Brueckner-Hartree-Fock (BHF) approach to calculate single-particle potentials of hyperons in dense nuclear matter.  The starting point of this approach is the use the NN, YN and YY potentials, supplemented by three-body forces. 

Whereas initial works on BHF computations were based on the J\"ulich and Nijmegen meson-exchange potentials, more recently the single-particle potentials for hyperons have been obtained with the $\chi$EFT approaches.  Within these schemes, the $\Lambda$ and $\Sigma$ single-particle potentials have been computed~\cite{Haidenbauer:2014uua, Petschauer:2015nea}.
The $\Sigma$-nuclear potential is found to be repulsive, whereas the $\Lambda$ single-particle potential is in good qualitative agreement with the empirical values extracted from hypernuclear data, becoming repulsive about two to three-times saturation density.  As for the $\Xi$ single-particle potential in nuclear matter,  values ranging between -$3$ to -$5$ MeV are determined, whereas the reported experimental value is larger.

The effect of three-body forces has been also studied, in particular for  the case of the $\Lambda$-nuclear interaction \cite{Haidenbauer:2016vfq}.  The inclusion of the three-body forces is important for obtaining binding energies of few nucleons, scattering observables and the nuclear saturation properties in non-relativistic schemes, such as BHF.  The implementation of three-body forces for the $\Lambda$-nuclear interaction in dense matter gives an extra repulsion at large densities, that could be relevant for the presence of hyperons in NSs  \cite{Logoteta:2019utx,Gerstung:2020ktv}, as we will discuss in the next section.

\subsubsection{Hyperons inside neutron stars: the hyperon puzzle}
\label{hyperonpuzzle}

As previously mentioned, a realistic scenario inside NSs involves the presence of neutrons and protons interacting, as well as electrons via weak equilibrium reactions when nucleons are involved\footnote{We have not explicitly mentioned the existence of muons, but those can  be also present when nucleons are considered.}. However, more exotic degrees of freedom could be also expected in the core of an NS.

\begin{figure*}[t]
\begin{center}
\resizebox{0.8\textwidth}{!}
{
\includegraphics[width=0.8\textwidth, angle=-90]{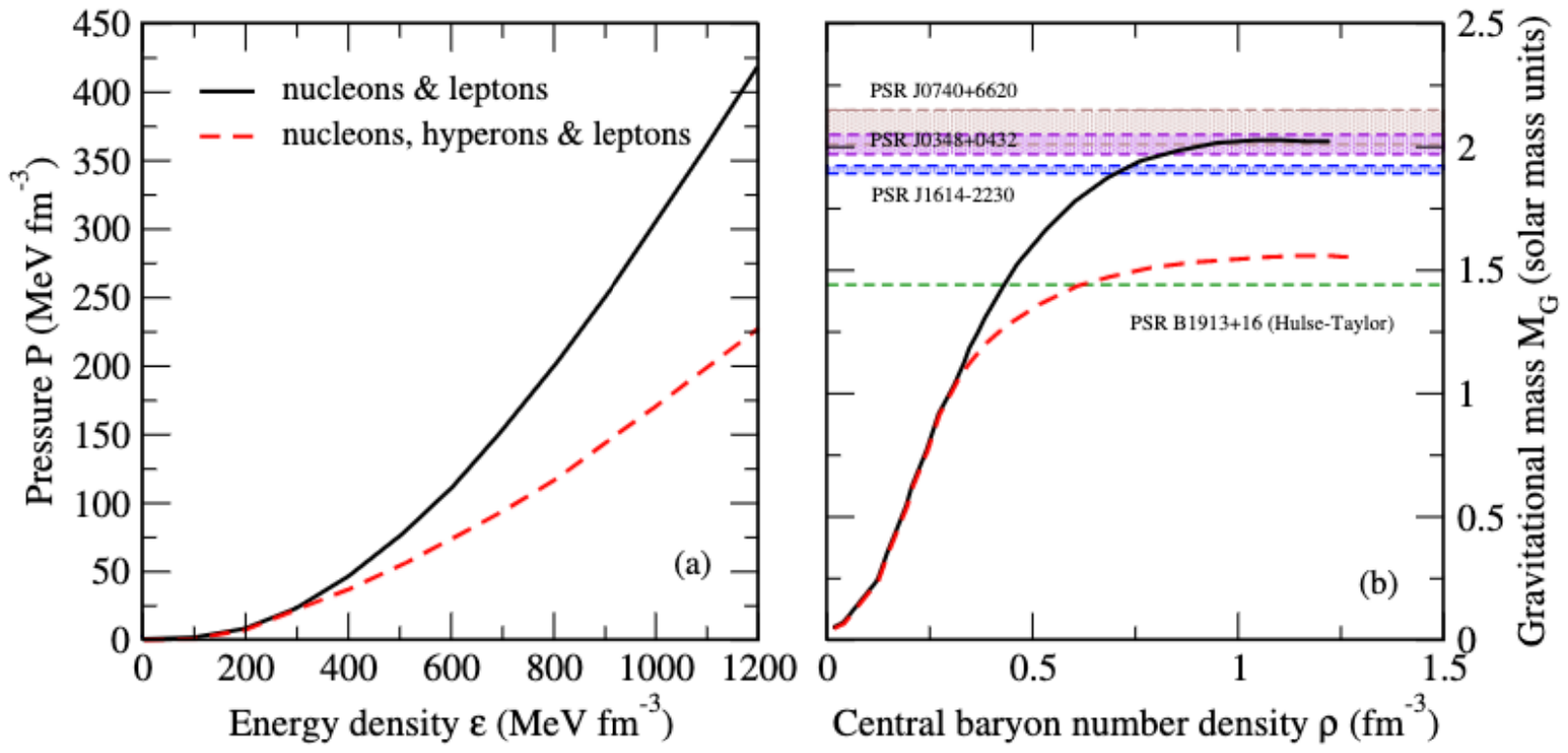}
}\caption{ The EoS (left panel) and the corresponding NS mass (right panel), without (black solid lines) and with (red dashed lines) hyperons. The mass of the Hulse-Taylor pulsar and the masses of  PSR J1614-2230 \cite{Demorest:2010bx,Fonseca:2016tux}, PSR J0348+0432 \cite{Antoniadis:2013pzd} and PSR J0740+6620 \cite{2019NatAs.tmp..439C}  are shown with horitzontal lines. Figure taken from Ref.~\cite{Vidana:2022tlx}.}
\label{fig:EOS}       
\end{center}
\end{figure*}

Hyperons in NSs were first taken into consideration in the seminal work of Ref.~\cite{1960SvA.....4..187A}. Ever since, the presence of hyperons in the interior of NSs have been thoroughly studied (for recent reviews see \cite{Tolos:2020aln,Burgio:2021vgk,Vidana:2018bdi,Chatterjee:2015pua}). Hyperons may appear inside NSs at densities of $\approx 2$-$3 \rho_0$. This is due to the fact that the nucleon chemical potential could be so large at these densities so that it is energetically more favourable to have hyperons than nucleons. As a result, the EoS becomes softer as the system relieves Fermi pressure, as observed in the left panel of Fig.~\ref{fig:EOS}. If the EoS becomes softer, then there is less pressure inside an NS, and, hence, the NS has less mass, as shown in the right panel of Fig.~\ref{fig:EOS}. The softening of the EoS may then lead to maximum masses not compatible with the $2M_{\odot}$ measurements, such as the masses of the PSR J1614-2230 \cite{Demorest:2010bx,Fonseca:2016tux}, PSR J0348+0432 \cite{Antoniadis:2013pzd} and PSR J0740+6620 \cite{2019NatAs.tmp..439C}, as seen in the right panel of Fig.~\ref{fig:EOS}. This fact is usually referred as the hyperon puzzle, and several solutions have been put forward in order to have hyperons in the interior of $2M_{\odot}$ NSs. Here we briefly comment on them. 

One solution to the hyperon puzzle takes into account stiff YN and YY interactions (see, for example, \cite{Bednarek:2011gd,Weissenborn:2011ut,Oertel:2014qza,Maslov:2015msa}). In this manner, the softening due to the presence of hyperons is overcome, thus reaching $2M_{\odot}$.   Another possible way of solving the puzzle is given by the stiffening of the EoS thanks to hyperonic three-body forces. The hyperonic three-body forces give an additional repulsion at large densities so the EoS becomes stiff enough in order to be able to obtain 2$M_{\odot}$ stars \cite{Haidenbauer:2016vfq,Logoteta:2019utx,Gerstung:2020ktv,Takatsuka2002,Takatsuka:2004ch,Vidana:2010ip,Yamamoto:2013ada,Yamamoto:2014jga,Lonardoni:2014bwa}. However, no general consensus has been reached. 
Other solutions consider the appearance of new species that could push the presence of hyperons  to larger densities, such as $\Delta$ baryons \cite{Schurhoff:2010ph,Drago:2014oja,Ribes:2019kno} or a kaon condensed phase (see discussion on kaon condensation in the next section). Moreover, solutions based on the appearance of non-hadronic degrees of freedom have been taken into account, such as an early phase transition to quark matter below the hyperon onset (see Refs.~\cite{Weissenborn:2011qu,Bonanno:2011ch,Klahn:2013kga,Lastowiecki:2011hh,Zdunik:2012dj} for recent papers).  And, finally, more exotic solutions  have been put forward, such as the use of modified gravity to accommodate hyperons inside $2M_{\odot}$ stars  \cite{Astashenok:2014pua}.

\subsection{Strange mesons: Antikaons}
\label{sec:kaoncond}

Up to now we have assumed that hadronic matter is made of baryons. However, another possible scenario inside NSs is the presence of bosonic matter,  in particular, the presence of strange mesons (antikaons denoted by $\bar K$)  in the core of NSs. In order to fully understand the plausibility of having strange mesons in the interior of NSs, we should first address the interaction of strange mesons with dense matter and how the properties of strange mesons are modified in a dense medium. 

Therefore, in this section we start by analysing the $\bar KN$ interaction and the role of the  $\Lambda(1405)$ resonance. Afterwards,  we address the interaction of strange mesons in a many-body system of nucleons. We continue by examining the role of strange mesons in HiCs, where a dense medium is produced. And, finally, we discuss the presence of antitkaons in NSs, and the phenomenon of kaon condensation.

\subsubsection{The $\bar KN$ interaction: the $\Lambda(1405)$}
\label{lambda1405}

The $\bar K N$ interaction  is governed by the presence of the $\Lambda(1405)$ state, which is a strange resonance with isospin $I=0$, spin and parity $J^P=1/2^-$ and strangeness $S=-1$. The $\Lambda(1405)$ was  predicted to be of molecular type more than 50 years ago by Dalitz and Tuan  \cite{Dalitz:1959dn,Dalitz:1959dq}. Since then, a lot of effort has been invested to understand its nature and, hence, the role of this state in the $\bar K N$ interaction. Several theoretical approaches have been used over the years, that include coupled-channel unitarized theories using meson-exchange models \cite{MuellerGroeling:1990cw,Haidenbauer:2010ch} or meson-baryon $\chi EFT$ \cite{Kaiser:1995eg, Oset:1997it,Oller:2000fj,Lutz:2001yb,GarciaRecio:2002td,Jido:2003cb,Borasoy:2005ie,Oller:2006jw,Feijoo:2018den}. Interestingly,  these works conclude that the dynamics of the $\Lambda(1405)$ is described by the superposition of two states, between the $\bar K N$ and $\pi \Sigma$ thresholds \cite{Oller:2000fj,Jido:2003cb,Hyodo:2007jq}, that can be seen experimentally in  reaction-dependent line shapes \cite{Jido:2003cb}.

\subsubsection{Antikaons in matter}
\label{matter}

Once we know the features of the $\bar KN$ interaction, we can then address the interaction of antikaons in a many-body system of nucleons. Over the last decades antikaons in nuclear matter have been extensively analyzed. The first works used relativistic mean-field models (RMF)  \cite{Schaffner:1996kv} or quark-meson coupling  schemes \cite{Tsushima:1997df} to obtain very large antikaon potentials of a few hundreds of MeVs at saturation density $\rho_0$. Nevertheless, the doubtful assumption of the low-density theorem led  these works to  determine such a large $\bar K$  potential in dense matter. The need of a description of the $\bar K N$ interaction taking into the $\Lambda(1405)$ in matter is essential.

One possible manner to proceed is by using unitarized theories in coupled channels in dense nuclear matter,  based on  $\chi EFT$  \cite{Waas:1996fy,Lutz:1997wt,Ramos:1999ku} or from meson-exchange schemes \cite{Tolos:2000fj,Tolos:2002ud}, in both cases including the strange degree of freedom. Different effects have to be taken into account to fully determine the behaviour of the $\Lambda(1405)$ in dense matter and, hence, the medium modified $\bar K N$ interaction. Those are: i) the implementation of Pauli blocking on baryons in the intermediate meson-baryon propagator \cite{Koch:1994mj}; ii) the inclusion of the $\bar K$ potential (or self-energy)  in the $\bar K$ propagation in dense matter \cite{Lutz:1997wt}; and  iii) the incorporation of self-energies of all  hadrons in the intermediate states \cite{Ramos:1999ku}.  Within these schemes, an attractive antikaon potential  is obtained  with values below 100 MeV at saturation density. Moreover,  the potential shows a considerable imaginary part, that is, the antikaon develops an important width because of the appearance of new decay channels of the antikaon in matter.

\subsubsection{Experiments and observations: heavy-ion collisions}
\label{hics}

A possible scenario to analyze the interaction of antikaons with a dense system and, hence, extract information on the antikaon potential is to study the creation and propagation of antikaons in HiCs for intermediate beam kinetic energies (GeV). This analysis,  however, requires the use of transport schemes to fully model the collisions.  Transport models can be understood as the link between the experiments and the physical processes, since they consider the production and propagation of all kind of species, such as strange mesons (see Ref.~\cite{Hartnack:2011cn} for a review on strangeness production). These schemes solve  semi-classical transport equations of the Boltzmann-type, coming from  non-equilibrium quantum field theory.

First transport calculations for antikaons in matter were performed neglecting the finite width of the antikaon potential in dense matter \cite{Cassing:1999es,Hartnack:2001zs}. Later on, the antikaon production was determined  by means of off-shell dynamics with full in-medium antikaon properties  within the Hadron-String-Dynamics (HSD) transport model \cite{Cassing:2003vz}. In this case, the  $\bar KN$ interaction in dense matter was obtained from the J\"ulich meson-exchange model \cite{Tolos:2000fj,Tolos:2002ud}. Recently,  strangeness production in HICs at (sub-)threshold energies of 1 - 2 AGeV based on the microscopic Parton-Hadron-String Dynamics (PHSD) transport approach has been studied, considering the in-medium antikaon properties from the $\chi EFT$ approach of Refs.~\cite{Tolos:2006ny,Tolos:2008di,Cabrera:2014lca}. Several experimental observables that involved strange mesons have been analyzed, such as rapidity distributions, $p_T$-spectra, the polar and azimuthal angular distributions, and directed and elliptic  flow in C$+$C, Ni$+$Ni, and Au$+$Au collisions. The comparison of this analysis with the experimental data from the KaoS, FOPI and HADES Collaborations lead to the conclusion that the modifications of the strange meson properties in dense nuclear matter are necessary to explain the data consistently \cite{Song:2020clw}.

\subsubsection{Experiments and observations: kaon condensation in neutron stars}
\label{kaoncondensation}

\begin{figure}[htb]
    \centering
    \includegraphics[width=0.43\textwidth, angle=-90]{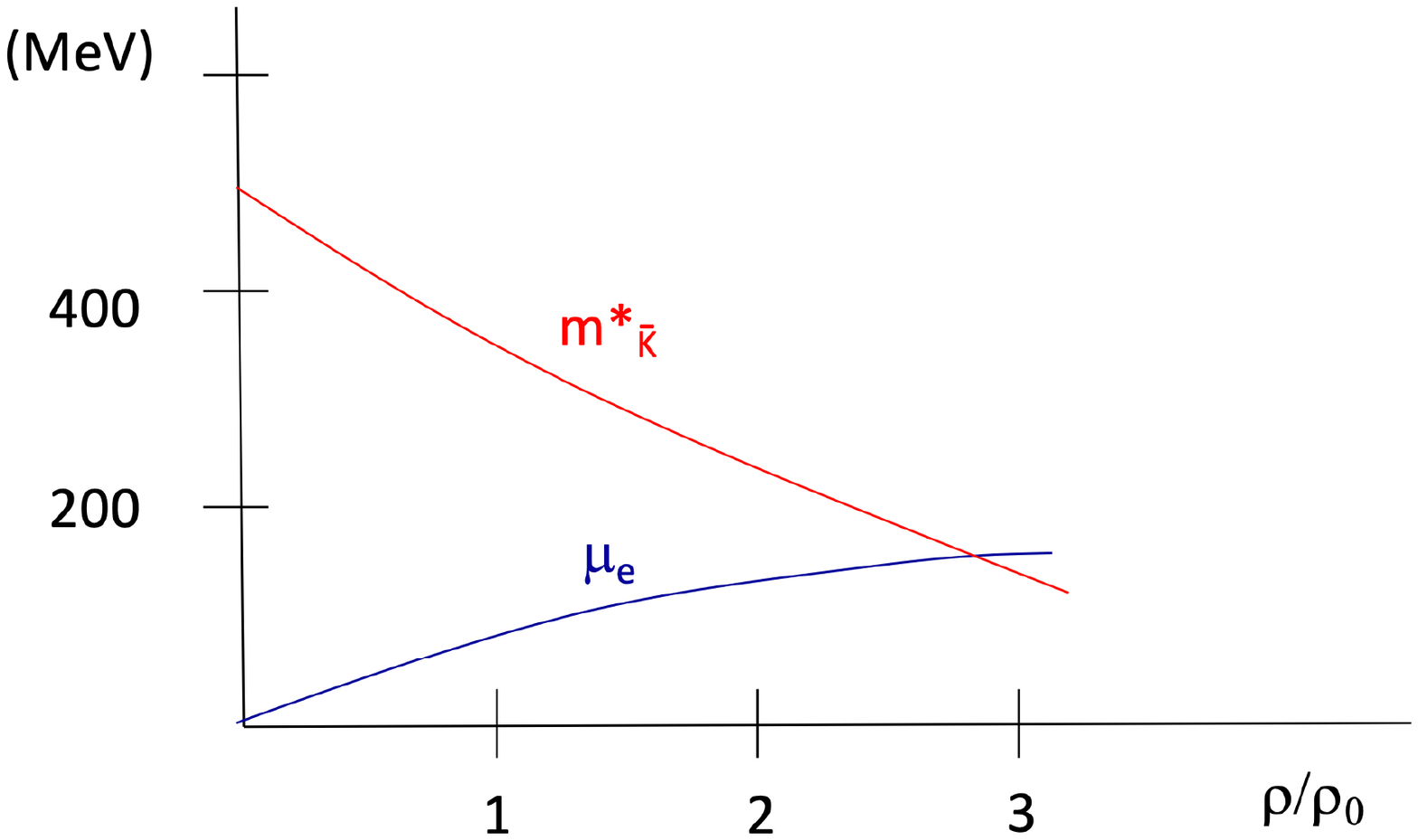}
    \caption{ Illustrative picture depicting the evolution of the electron chemical potential $\mu_e$ and the antikaon effective mass $m^*_{\bar K}$ with baryon density in the interior of NSs. Taken from Ref.~\cite{Tolos:2020aln}.}
    \label{fig:kaoncond}
\end{figure}

The presence of antikaons  is another possible scenario inside the core of NSs. As mentioned earlier, the composition of matter in NSs is found by demanding equilibrium against weak interaction processes. In particular, for matter composed of neutrons, protons and electrons, the weak interaction transitions are given by
\begin{eqnarray}
n \rightarrow p \  e^- \ \bar  \nu_e \nonumber \\
e^- \ p \rightarrow n \  \nu_e ,
\end{eqnarray} 
with $\mu_n=\mu_p+\mu_e$ and $\rho_p=\rho_e$, with $\rho=\rho_p+\rho_n$. However, if the chemical potential of the electron substantially increases with density in the interior of an NS, antikaons might be produced instead of electrons as the following weak reactions could become energetically more favourable  
\begin{eqnarray}
n \leftrightarrow p + \bar{K} .
\end{eqnarray}
In order for these reactions to take place, the chemical potential of the electron for a given density in the core of an NS should be larger than the effective mass of antikaons at that density, that means, $\mu_e > m^*_{\bar K}$. If this is the case,  the phenomenon of kaon condensation would take place  as antikaons would appear and form a condensed medium.

The possible existence of kaon condensation in NSs was considered in the pioneering work of Ref.~\cite{Kaplan:1986yq}.  The question that needs to be answered is whether the mass of antikaons could be largely modified in  the nuclear medium. This is the case of some phenomenological schemes, in particular those based in RMF models (see, for example,  the recent results in Refs.~\cite{Gupta:2013sna,Thapa:2021kfo,Malik:2021nas,Muto:2021jms}). However, the large modification in the mass of antikaons embedded in a nuclear medium is not obtained in  microscopic unitarized schemes (see, for example, Refs.~\cite{Lutz:1997wt,Ramos:1999ku,Tolos:2000fj,Tolos:2002ud,Tolos:2006ny,Tolos:2008di,Cabrera:2014lca,Lutz:2007bh}).

\begin{figure} [t]
    \centering
    \includegraphics[width=0.85\textwidth]{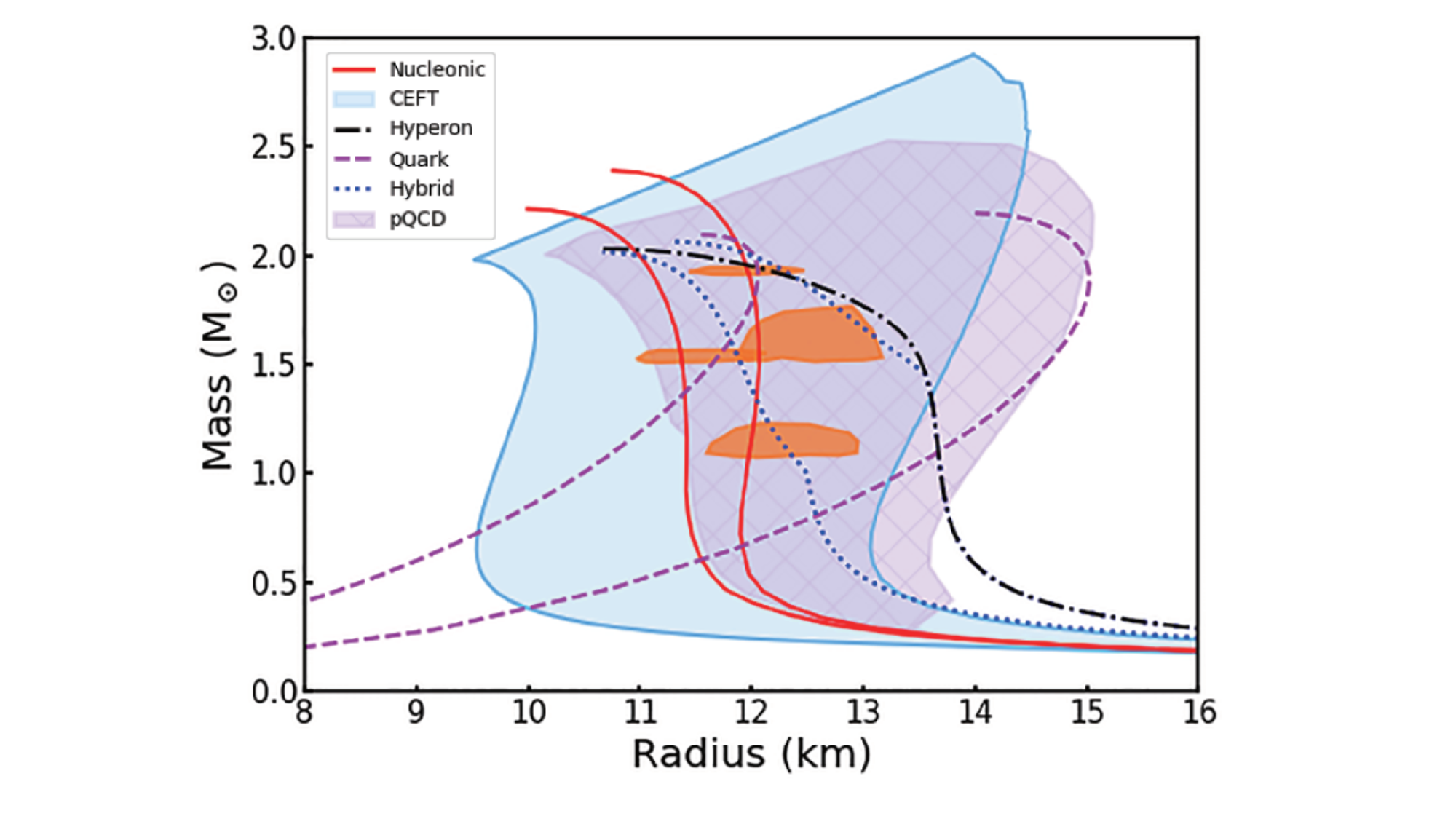}
    \caption{ Constraints from pulse profile modelling of rotation-powered pulsars with eXTP. Figure adapted from Ref.~\cite{Watts:2018iom} and taken from Ref.~\cite{Tolos:2020aln}. }
    \label{fig:mass-radius}
\end{figure}

\section{Conclusions and Outlook}

In this lecture we have considered the properties of strange hadronic matter in a dense medium and, more precisely, inside NSs. In particular, we have discussed two possible scenarios in the interior of NSs, that is, the presence of hyperons which might lead to the hyperon puzzle and the phenomenon of kaon condensation.

To finalize we would  like to discuss the future venue to address strange matter inside NSs through X-ray observations. In Fig.~\ref{fig:mass-radius} we show the mass-radius diagram for NSs taking into account different possible scenarios inside NSs, together with constraints from pulse profile modelling with eXTP \cite{Watts:2018iom}. The expected constraints from pulse profile modelling of rotation-powered pulsars with eXTP are shown with the orange error contours for PSR J1614-2230 \cite{Demorest:2010bx,Fonseca:2016tux}, PSR J2222-0137 \cite{Kaplan:2014mka}, PSRJ0751+1807 \cite{Desvignes:2016yex} and PSR J1909-3744 \cite{Desvignes:2016yex}), whose masses are known precisely. The EoS models\footnote{We note the existence of the online service CompOSE repository that provides data tables for different state of the art EoSs ready for use in astrophysical applications, nuclear physics and beyond \cite{compose,CompOSECoreTeam:2022ddl,Dexheimer:2022qhn}. }  include nucleons (models AP3 and AP4) \cite{Akmal:1998cf},  quarks (u,d,s quarks) \cite{Li:2016khf,Bhattacharyya:2016kte}, nucleons and hyperons (inner core with nucleons and hyperons, outer core with only nucleons) \cite{Bednarek:2011gd}, or quarks and nucleons giving rise to hybrid stars (inner core of  quarks, outer core of nucleons)\cite{Zdunik:2012dj}. The CEFT region shows the range of the nucleonic $\chi$EFT EoS \cite{Hebeler:2013nza}, while the pQCD domain results from interpolating CEFT at low densities and matching to perturbative QCD (pQCD) computations at higher densities \cite{2014ApJ...789..127K}. 

From this figure it is clear the need of having precise simultaneous mass-radius observations to disentagle between the theoretical predictions for different types of dense matter inside NSs. Nonetheless, other observations are very much welcome, such as those coming from gravitational wave emission of NS binary mergers.

\section*{Acknowledgments}

L.T. acknowledges support from CEX2020-001058-M (Unidad de Excelencia ``Mar\'{\i}a de Maeztu"), PID2019-110165GB-I00 and PID2022-139427NB-I00 financed by the Spanish MCIN/AEI/10.13039/501100011033/FEDER, UE as well as from the Generalitat de Catalunya under contract 2021 SGR 171,  by  the EU STRONG-2020 project, under the program  H2020-INFRAIA-2018-1 grant agreement no. 824093, and by the CRC-TR 211 'Strong-interaction matter under extreme conditions'- project Nr. 315477589 - TRR 211.

\bibliographystyle{unsrt}
\bibliography{tolos_Zakopane_rev.bib}


\end{document}